\documentclass[journal]{IEEEtran}

\usepackage{cite}
\usepackage{amsmath,amssymb,amsfonts}
\usepackage{algorithmicx}
\usepackage{graphicx} 
\usepackage{textcomp}
\usepackage{wrapfig,colortbl}
\usepackage{algpseudocode}
\usepackage{xcolor}
\usepackage{tikz}
\usepackage{tikz}
\usetikzlibrary{calc}
\usetikzlibrary{matrix, positioning, shapes, fit}
\usepackage{caption}
\usepackage{tabularx}
\usepackage{adjustbox}
\usepackage{booktabs} 
\usepackage{makecell}
\usepackage{array}
\usepackage{soul}
\usepackage{dblfloatfix}

\usepackage[colorlinks=true, linkcolor=blue, citecolor=blue, urlcolor=blue]{hyperref}

\def\BibTeX{{\rm B\kern-.05em{\sc i\kern-.025em b}\kern-.08em
 T\kern-.1667em\lower.7ex\hbox{E}\kern-.125emX}}

\definecolor{PastelBlue}{RGB}{173, 216, 230} 
\definecolor{PastelPink}{RGB}{255, 182, 193}
\definecolor{RoyalBlue}{RGB}{30,77,216}
\definecolor{SunsetOrange}{RGB}{255,107,26}
\definecolor{TurquoiseGreen}{RGB}{0,194,160}
\definecolor{GoldenYellow}{RGB}{255,217,61}
\definecolor{FuchsiaPink}{RGB}{230,57,143}
\definecolor{myblue}{HTML}{1F77B4}
\definecolor{myorange}{HTML}{FF7F0E}
\definecolor{myteal}{HTML}{2CA02C}
\definecolor{myyellow}{HTML}{FFD700}
\definecolor{mymagenta}{HTML}{D62728}

\newcolumntype{Y}{>{\centering\arraybackslash}X}

\tikzset{
 matrixcell/.style={
 draw=black!60, minimum width=8mm, minimum height=8mm, anchor=center, font=\footnotesize
 },
 framedblock/.style={
 rounded corners=2pt, line width=1.2pt
 },
 labelbox/.style={
 draw=black, fill=white, inner sep=3pt, font=\small
 }
}

\ifCLASSINFOpdf
\else
 \usepackage[dvips]{graphicx}
\fi
\usepackage{url}

\usepackage{graphicx}

\begin{document}

\title{Efficient Convolutional Forward Model for Passive Acoustic Mapping and Temporal Monitoring}

\author{Tatiana Gelvez-Barrera,
Barbara Nicolas, Bruno Gilles, Adrian Basarab, Denis Kouamé
\thanks{This work was supported by LABEX CELYA (ANR-10- LABX-0060) and LABEX PRIMES (ANR-11-LABX-0063) of Université de Lyon, within the program ``Investissements d'Avenir" (ANR-11-IDEX-0007) operated by the French National Research Agency (ANR), and the ``CAVIIAR" Project (ANR-22-CE19-0006), operated by the French National Research Agency (ANR).}
\thanks{T. Gelvez-Barrera, B. Nicolas and A. Basarab are with Université Claude Bernard Lyon 1, INSA-Lyon, CNRS, Inserm, CREATIS UMR5220, U1294,
F-69100, Lyon, France. (e-mail: tatiana.gelvez@saber.uis.edu.co, barbara.nicolas@creatis.insa-lyon.fr, adrian.basarab@creatis.insa-lyon.fr). }
\thanks{B. Gilles is with LabTau, Inserm, U1032,
Université Claude Bernard Lyon 1, F-69003, Lyon, France. (e-mail: bruno.gilles@inserm.fr).}
\thanks{T. Gelvez-Barrera and D. Koumé are with IRIT, Université de Toulouse, CNRS, Toulouse, France. (e-mail: denis.kouame@irit.fr).}
}

\markboth{Journal of \LaTeX\ Class Files, Vol. 14, No. 8, August 2015}
{Shell \MakeLowercase{\textit{et al.}}: Bare Demo of IEEEtran.cls for IEEE Journals}
\maketitle

\begin{abstract}
Passive acoustic mapping (PAM) is a key imaging technique for characterizing cavitation activity in therapeutic ultrasound applications. Recent model-based beamforming algorithms offer high reconstruction quality and strong physical interpretability. However, their computational burden and limited temporal resolution restrict their use in applications with time-evolving cavitation. To address these challenges, we introduce a PAM beamforming framework based on a novel convolutional formulation in the time domain, which enables efficient computation. In this framework, PAM is formulated as an inverse problem in which the forward operator maps spatiotemporal cavitation activity to recorded radio-frequency signals accounting for time-of-flight delays defined by the acquisition geometry. We then formulate a regularized inversion algorithm that incorporates prior knowledge on cavitation activity. Experimental results demonstrate that our framework outperforms classical beamforming methods, providing higher temporal resolution than frequency-domain techniques while substantially reducing computational burden compared with iterative time-domain formulations.
\end{abstract}


\IEEEpeerreviewmaketitle

\section{Introduction}

\IEEEPARstart{T}{herapeutic} ultrasound procedures such as high-intensity focused ultrasound (HIFU), histotripsy, or ultrasound-enhanced drug delivery can induce cavitation activity through rapid pressure fluctuations, producing gas or vapor microbubbles that emit acoustic signals~\cite{moonen2025focused,denner2023modeling}. Depending on the acoustic pressure and medium properties, microbubbles may undergo stable oscillations (non-inertial cavitation) or collapse violently within microseconds (inertial cavitation)~\cite{tan2024modelling, coussios2008applications}. Each cavitation regime is closely linked to therapeutic outcomes, bio-effects, or pathological conditions~\cite{saletesresearch}. Consequently, monitoring cavitation activity is essential for ensuring safety and efficacy during ultrasound-based therapies~\cite{chen2016dynamic}.

Passive Acoustic Mapping (PAM) characterizes the spatial and temporal distribution of cavitation activity by beamforming radio-frequency (RF) signals recorded on a sensor array. Early PAM methods adapted classical acoustic localization~\cite{gyongy2009passive} and advanced techniques such as robust beamforming~\cite{coviello2015passive}, coherence-based weighting~\cite{boulos2018weighting}, and compressive sensing~\cite{crake2018passive} to estimate cavitation maps. More recently, PAM beamforming has been formulated as an inverse problem, enhancing interpretability and enabling the integration of prior knowledge on cavitation activity through regularization~\cite{lachambre2024inverse, GelvezBarrera2025}.

Most of such PAM methods build upon classical beamforming strategies,
which process RF data either in time domain (TD) or in frequency-domain (FD).
TD methods offer high temporal resolution and straightforward implementation, but generally provide lower spatial resolution and higher computational cost, especially in iterative variational approaches~\cite{GelvezBarrera2025}. In contrast, FD methods, often based on Cross Spectral Matrix (CSM) analysis~\cite{polichetti2020use, sivadon2020pisarenko}, leverage spectral information to achieve improved spatial resolution, greater computational efficiency, and better discrimination of cavitation regimes. However, they require long signal segments, which can reduce temporal tracking capabilities, and assume signal stationarity, which may not hold for rapidly evolving cavitation events.

To overcome such limitations, this work proposes an efficient convolutional beamforming framework for PAM. More precisely, our approach, hereafter
referred to as Time-Domain Convolutional Model-based Passive
Acoustic Mapping (TD-CM-PAM), formulates PAM as
a time-domain inverse problem, in which the forward model
relates spatiotemporal cavitation source activity to the recorded
RF signals by capturing time-of-flight delays imposed by the
acquisition geometry.
As a novel contribution, this forward operator is expressed as a linear convolution, enabling efficient implementation. Based on this formulation, a regularized inversion algorithm is applied, leveraging prior information of cavitation activity.

Experiments on scenarios with evolving cavitation activity demonstrate that the proposed TD-CM-PAM outperforms standard TD and FD beamforming methods, providing higher temporal resolution than FD approaches while significantly reducing computational cost compared to iterative TD methods.

\section{Convolutional Beamforming Model for PAM}

In what follows, $(x, y, z)$ refer to the lateral, azimuthal, and axial spatial dimensions, respectively, and $(t)$ to the temporal dimension.
Let $\mathbf{Y} \in \mathbb{R}^{N_m \times N_t}$ be the matrix containing the RF signals recorded by an array with $N_m$ sensors, sampled at $N_t$ temporal samples with a sampling frequency $f_s$, so that, the $k^\text{th}$ temporal sample refers to the time instant $t_k = k / f_s$.


The RF signals can be expressed in vectorized form as:
\begin{equation}
\mathbf{y} \in \mathbb{R}^{N_m N_t} =
\begin{bmatrix}
\mathbf{y}_1^\top & \ldots & \mathbf{y}_m^\top & \ldots & \mathbf{y}_{N_m}^\top
\end{bmatrix}^\top,
\end{equation}
where $\mathbf{y}_m \in \mathbb{R}^{N_t}$ denotes the vector form of the signal recorded by the sensor located at position $\vec{\mathbf{r}}_m = (x_m, 0, z_m)$, for $m = 1, \ldots N_m$. 

Let $\boldsymbol{\mathcal{X}} \in \mathbb{R}^{N_x \times N_z \times N_t}$ denote the cavitation spatiotemporal datacube, parameterized by the lateral–axial coordinates $(x, z)$ at a fixed azimuthal position $y_0$ and evolving over time $t$. The vector $\boldsymbol{\mathcal{X}}_{i, j, :} \in \mathbb{R}^{N_t}$ represents the temporal waveform at spatial location $(x_i, z_j)$, with $i = 1, \dots, N_x$ and $j = 1, \dots, N_z$ indexing lateral and axial pixels, respectively.

Using a flattened spatial index $n \in \{1, \ldots, N\}$ associated to a coordinate tuple $(i_n, j_n)$, with $N = N_x N_z$, the datacube can be vectorized as
\begin{equation}
 \mathbf{x} \in \mathbb{R}^{N N_t} =
 \begin{bmatrix}
 \mathbf{x}_1^\top & \cdots & \mathbf{x}_n^\top & \cdots & \mathbf{x}_N^\top
 \end{bmatrix}^\top,
\end{equation}
where $\mathbf{x}_n \in \mathbb{R}^{N_t}$ is the vector form of the temporal waveform $\boldsymbol{\mathcal{X}}_{i_n, j_n, :}$ located at position $\vec{\mathbf{r}}_n = (x_n, y_0, z_n)$.

Model-based inversion methods rely on a forward model that describes the physical relationship between the observed data and the underlying source distribution. Hence, beamforming for PAM can be formulated as a model-based inverse problem, assuming the following linear forward model:
\begin{equation}
 \mathbf{y} = \mathbf{A} \mathbf{x} + \boldsymbol{\eta},
 \label{eq:forward_model}
\end{equation}
where $\mathbf{A} \in \mathbb{R}^{N_m N_t \times N N_t}$ denotes a linear forward operator, 
and $\boldsymbol{\eta} \in \mathbb{R}^{N_m N_t}$ accounts for additive acquisition noise.

\subsection{Convolutional Forward Model}
\label{subsubsec:convolutionforward}
The operator $\mathbf{A}$ was modeled as a sparse binary matrix in~\cite{GelvezBarrera2025}, uniquely based on the acoustic time-of-flight defined by the acquisition geometry, where the discrete propagation delay $\delta_{m,n}$ between sensor $m$ and pixel $n$ is given by:
\begin{equation}
\delta_{m,n} := \left\lfloor \frac{\lVert \vec{\mathbf{r}}_m - \vec{\mathbf{r}}_n \rVert_2}{c} \cdot f_s \right\rceil,
\label{eq:delay_definition}
\end{equation}
where $c$ denotes the speed of sound in the medium, assumed constant and known, and $\lfloor \cdot \rceil$ represents the rounding operator to the nearest integer, ensuring that the propagating wavefront aligns with a discrete temporal sample.

For certain configurations, this operator may exhibit a block-Toeplitz structure, enabling a convolutional formulation. To our knowledge, such a formulation has never been introduced for PAM and may offer a significant reduction in computational cost. Accordingly, this subsection presents the forward model expressed in terms of convolutions with binary kernels.

The block-Toeplitz structure arises when the sensor pitch, $d = |x_{m+1} - x_{m}|, \forall m$ matches the lateral pixel width in the image, i.e., $d = |x_{i+1} - x_{i}|, \forall i$, since such sensor–pitch to pixel–width matching produces symmetry and shift-invariance.

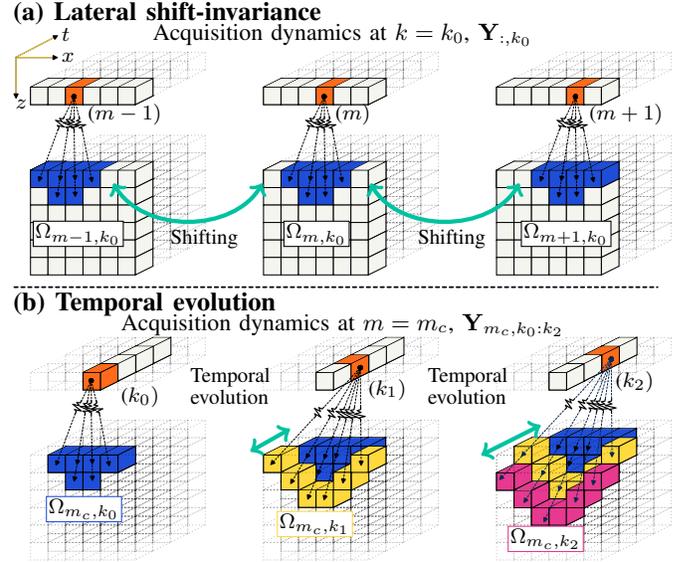
\begin{figure}[tb!]
    \centering
    \begin{adjustbox}{width=1\linewidth}
        \begin{tikzpicture}
            \node[anchor=north west, inner sep=0] (img) 
                at (0,0) {\includegraphics[width=1\linewidth]{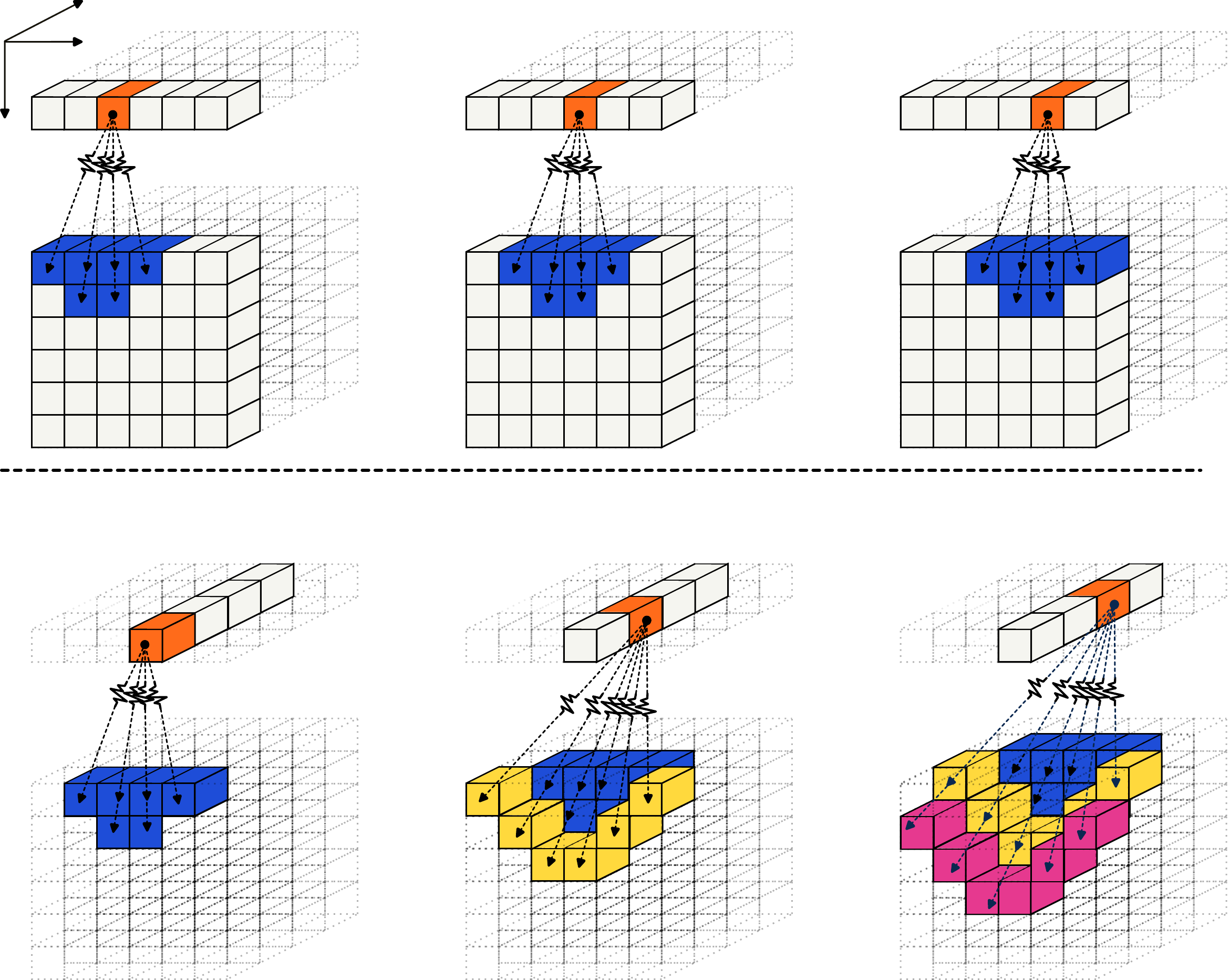}};
        

\node[anchor = north west, inner sep=0] at (0\linewidth,0.45) {\normalsize{\textbf{(a) Lateral shift-invariance}}};

\node [anchor = center] at (0.5\linewidth,0.0) {\small{Acquisition dynamics at $k = k_0$, $\mathbf{Y}_{:, k_0}$}};

\node [anchor = north west] at (0.06\linewidth,-0.12) {\footnotesize{$x$}};
\node [anchor = north west] at (-0.01\linewidth,-0.75) {\footnotesize{$z$}};
\node [anchor = north west] at (0.06\linewidth,0.21) {\footnotesize{$t$}};

\node [anchor = north west] at (0.1\linewidth,-0.8) {\footnotesize{(${m-1})$}};

\node [anchor = north west, fill = white, draw = black, inner sep = 0.08em] at (0.03\linewidth,-2.5) {\footnotesize{${{\Omega}}_{m-1,k_0}$}};

\node [anchor = north west] at (0.475\linewidth,-0.8) {\footnotesize{(${m})$}};

\node [anchor = north west, fill = white, draw = black, inner sep = 0.08em] at (0.41\linewidth,-2.5) {\footnotesize{${{\Omega}}_{m,k_0}$}};

\node [anchor = north west] at (0.86\linewidth,-0.8) {\footnotesize{(${m+1})$}};

\node [anchor = north west, fill = white, draw = black, inner sep = 0.08em] at (0.77\linewidth,-2.5) {\footnotesize{${{\Omega}}_{m +1,k_0}$}};

\draw[anchor=north west, <->, ultra thick, color={rgb,255:red,0; green,194; blue,160}](1.35, {-2}) to[out={-60}, in={-120}] (3.35, {-2});
\node [anchor=north west] at (0.225\linewidth, {-2.5}) {\footnotesize{Shifting}};

\draw[ anchor = north west, <->, ultra thick, color={rgb,255:red,0; green,194; blue,160}]
(4.8, -2) to[out=-60, in=-120] (6.8, -2);

\node [anchor = north west] at (0.6\linewidth,-2.5) {\footnotesize{Shifting}};

\node[anchor = north west, inner sep=0] at (0\linewidth,-3.45) {\normalsize{\textbf{(b) Temporal evolution}}};

\node [anchor = center] at (0.5\linewidth,-3.9) {\small{Acquisition dynamics at $m = m_c$, $\mathbf{Y}_{m_c, k_0:k_2}$}};

\node [anchor = north west] at (0.15\linewidth,-4.6) {\footnotesize{(${k_0})$}};

\node [anchor = north west, fill = white, draw = RoyalBlue, inner sep = 0.08em] at (0.05\linewidth,-6.2) {\footnotesize{${{\Omega}}_{m_c,k_0}$}};

\node [anchor = north west] at (0.525\linewidth,-4.5) {\footnotesize{(${k_1})$}};

\node [anchor = north west, fill = white, draw = GoldenYellow, inner sep = 0.08em] at (0.4\linewidth,-6.4) {\footnotesize{${{\Omega}}_{m_c,k_1}$}};

\node [anchor = north west] at (0.9\linewidth,-4.4) {\footnotesize{(${k_2})$}};

\node [anchor = north west, fill = white, draw = FuchsiaPink, inner sep = 0.08em] at (0.75\linewidth,-6.6) {\footnotesize{${{\Omega}}_{m_c,k_2}$}};

\draw[ anchor = north west, <->, ultra thick, color={rgb,255:red,0; green,194; blue,160}]
(0.36\linewidth, -5.60) to (0.42\linewidth, -5.3);
 \node [anchor = north west,  rotate=0] at (0.255\linewidth,-4.35) {\footnotesize{Temporal}};

  \node [anchor = north west,  rotate=0] at (0.255\linewidth,-4.65) {\footnotesize{evolution}};

\draw[ anchor = north west, <->, ultra thick, color={rgb,255:red,0; green,194; blue,160}]
(0.71\linewidth, -5.65) to (0.8\linewidth, -5.25);
 \node [anchor = north west, rotate =0] at (0.615\linewidth,-4.35) {\footnotesize{Temporal}};
  \node [anchor = north west, rotate =0] at (0.615\linewidth,-4.65) {\footnotesize{evolution}};

\end{tikzpicture}
\end{adjustbox}
 \caption{Passive acquisition dynamics under the sensor–pitch to pixel–width matching configuration. (a) For a single temporal sample, the contributors set $\Omega_{m,k}$ forms a paraboloid-like delay pattern. Neighboring sensors exhibit identical patterns differing only by a lateral shift. (b) As time progresses, the delay pattern expands outward in the $(x,z)$ plane, so more distant pixels may contribute. Thus, patterns evolve over time, and recorded signals result from the cumulative superposition of contributions across successive temporal slices.}
\label{fig:kernel_structure}
\end{figure}

Let $\boldsymbol{\mathcal{K}} \in {\{ 0,1 \}}^{N_w \times N_z \times N_t}$ be a 3D convolution kernel, where
$N_w = N_x + 2 \left\lceil N_m/2 \right\rceil$. This kernel models the delay effects across the lateral, axial, and temporal dimensions. The extended lateral size $N_w$ ensures that the kernel fully captures the maximum possible lateral shifts in the array. The kernel structure $\boldsymbol{\mathcal{K}}$ is defined by analyzing the passive acquisition dynamics under the sensor–pitch to pixel–width matching configuration detailed below.

\subsubsection{\textit{\textbf{Lateral shift-invariance}}} 
Figure~\ref{fig:kernel_structure}~(a) depicts the passive acquisition process in the linear probe array, considering a single temporal sample $k = k_0$. 
Due to the nature of acoustic wave propagation, there exists a set of spatial positions 
whose emissions arrive simultaneously at sensor $m$ at time $t_{k_0}$. 
This set of potential contributors, denoted by $\Omega_{m,k_0}$, is given by:
\begin{equation}
 n \in \Omega_{m,k_0} \quad \text{if} \quad \delta_{m,n} = k_0.
\end{equation}
Visually, $\Omega_{m,k_0}$ forms a paraboloid-like delay pattern in the spatial domain, as illustrated by the blue pixels in Fig.\ref{fig:kernel_structure}~(a). The RF signal at each sensor $m$ is then the superposition of sources at the pixels in the corresponding $\Omega_{m,k_0}$ sets.

When the sensor-pitch matches the pixel-width, neighboring sensors exhibit identical delay patterns, differing only by a lateral shift. Hence, for any sensor $m$, the set $\Omega_{m,k_0}$ can be obtained by laterally shifting the set $\Omega_{m_c,k_0}$ of the central sensor located at $m_c = \lceil N_m/2 \rceil$, where $\lceil \cdot \rceil$ maps a real number to the least integer not smaller than it. The RF signals across all sensors can be then computed by the dot product of the plane image at time $k_0$ with the laterally shifted central kernel, containing ones at the positions specified by $\Omega_{m_c,k_0}$.

\subsubsection{\textbf{\textit{Temporal evolution}}} As the time index $k$ increases, the resulting delay pattern expands outward in the $(x,z)$ plane. 
In other words, as time progresses, the potential contributors can be located farther from the sensors. 
As a result, the patterns evolve over time, and the structure of the $k^{\text{th}}$ kernel slice $\boldsymbol{\mathcal{K}}_{:,:,k}$ is given by the positions in the sets $\Omega_{m_c,k}, \forall k$.

Taking into account this temporal evolution, the passive acquisition results in a cumulative sum, where the recorded signal at sensor $m$ and temporal sample $k$, $\mathbf{Y}_{m,k}$, is determined by the current kernel $\boldsymbol{\mathcal{K}}_{:,:,k}$ and previous temporal kernels, capturing earlier emissions from farther positions, as shown in Fig.~\ref{fig:kernel_structure}~(b).

Based on these dynamics, the passive acquisition process can be formulated through a \emph{full}-mode convolution, $(*)$, between the datacube $\boldsymbol{\mathcal{X}}$ and the kernel $\boldsymbol{\mathcal{K}}$. The resulting tensor
$\boldsymbol{\mathcal{Y}} \in \mathbb{R}^{(N_x + N_w - 1) \times (2N_z - 1) \times (2N_t - 1)}$
contains the contributions of all delayed wavefronts, where the recorded RF signal, $\mathbf{Y}$, corresponds to the axial slice at index $N_z$ of $\boldsymbol{\mathcal{Y}}$. This single slice can be expressed as a sum of 2D convolutions:
\begin{equation}
\boldsymbol{\mathcal{Y}}_{:,\, N_z,\, :} = \sum_{j=1}^{N_z} \boldsymbol{\mathcal{K}}_{:,\, j,\, :} * \boldsymbol{\mathcal{X}}_{:,\, N_z - j + 1,\, :}.
\label{eq:sumConvtwo-dimensional}
\end{equation}

Given that $\boldsymbol{\mathcal{Y}}_{:,N_z,:}$ contains contributions beyond the actual measurement grid, the observed signal $\mathbf{Y}$ is obtained by extracting a specific lateral-temporal region from $\boldsymbol{\mathcal{Y}}_{:, N_z,:}$. We formalize this extraction by using the operator $\mathcal{E}(\cdot)$, which isolates the subset of indices
from $N_x + 1$ to $N_x + 2\left\lceil N_m/2 \right\rceil$ in the lateral dimension 
and from $1$ to $N_t$ in the temporal dimension. 
Thus, finally the PAM forward model can be written in terms of convolution operations as:
\begin{equation}
\mathbf{Y} = \mathcal{E}\left( \sum_{j=1}^{N_z}   \boldsymbol{\mathcal{K}}_{:,\, j,\, :} * \boldsymbol{\mathcal{X}}_{:,\, N_z - j,\, :} \right) + \mathbf{W}, 
\label{eq:ConvFormulation}
\end{equation}
where $\mathbf{W} \in \mathbb{R}^{N_m \times N_t}$ represents additive acquisition noise.

\subsubsection{Computational complexity analysis}
The computational complexity of the forward model, when expressed as the matrix–vector product $\mathbf{y} = \mathbf{A}\mathbf{x}$, can be written as
\[
\mathcal{O}\big((N_m N_t)\,(N N_t)\big)
= \mathcal{O}\big(N_m\,N\,N_t^{2}\big).
\]
In practice, $\mathbf{A}$ is highly sparse, and it can be implemented in a memory-efficient manner via function-based operators, avoiding the need to explicitly store the full matrix. Nonetheless, the time complexity remains high.

Meanwhile, the computational complexity of the convolutional forward model in Eq.~\eqref{eq:ConvFormulation}, is determined by computing $N_z$ full 2D convolutions between an $N_x\times N_t$ slice and an $N_w\times N_t$ kernel, which requires
\[
\mathcal{O}\big(N_z (N_x N_t) (N_w N_t)\big) = \mathcal{O}\big(N_w N N_t^2\big),
\]
where $N_w = N_x + 2 \lceil N_m/2 \rceil$ is bounded by $N_x + N_m$, so the cost scales similarly to the matrix–vector formulation.

This formulation can be implemented efficiently via the FFT, where each 2D full-convolution output has size $(N_x + N_w - 1) \times (2N_t - 1)$. The computational cost then becomes:
\[
\mathcal{O}\big(N_z (N_x + N_w) (2 N_t) \log(( N_x + N_w) (2 N_t))\big),
\]
which reduces the temporal–lateral complexity from quadratic to quasi-linear, making it less expensive than the explicit matrix–vector formulation. 

\subsection{Regularized Inversion for Passive Acoustic Mapping}
\label{subsec:inverse}

The proposed forward model with the convolutional operator introduced in Sec.~\ref{subsubsec:convolutionforward} serves as the foundation for our TD-CM-PAM framework via regularized inversion strategies. Without loss of generality, we formulate the inverse problem using vector notation, regardless of the fact that the acquisition corresponds to a convolution, since it is still a linear operation. The reconstruction of the spatiotemporal cavitation activity signal, $\mathbf{x}$, is then expressed as the following inverse problem:
\begin{equation}
\underset{\mathbf{x} \in \mathbb{R}^{N N_t}}{\text{minimize}}
\quad
\frac{1}{2} | \mathbf{y} - \mathbf{A} \mathbf{x} |_2^2 + \lambda \mathcal{R}(\mathbf{x}),
\end{equation}
where the first term represents the data fidelity term, ensuring agreement with the recorded RF signals, and the second term is the regularization function, encoding prior knowledge about the structure of the cavitation activity signal to address the ill-posedness of the inverse problem. The parameter $\lambda$ balances the influence of the regularization term.

Based on the study presented in~\cite{GelvezBarrera2025}, we choose a joint regularizer that combines sparsity (Sp) with Regularization by Denoising (ReD) to guide the reconstruction process, providing improved localization accuracy, enhanced shape identification, and effective noise suppression. This approach, denoted as TD-CM-PAM$_\text{SpReD}$, is formulated as:

\begin{equation}
\hat{\mathbf{x}} \in \underset{\mathbf{x} \in \mathbb{R}^{N N_t}}{\arg\min}
\left\{ 
\frac{1}{2} \left\| \mathbf{y} - \mathbf{A} \mathbf{x} \right\|_2^2 
+ \lambda \left\| \mathbf{x} \right\|_1 
+ \mu \mathcal{R}_{\mathrm{D}}(\mathbf{x})
\right\},
\label{eq:inverseProblem}
\end{equation}
where, $\|\mathbf{x}\|_1 = \sum_{i,j,k} |\mathcal{X}_{i,j,k}|$ denotes the $\ell_1$-norm promoting sparsity, and 
$\mathcal{R}_{\mathrm{D}}(\mathbf{x}) = \frac{1}{2} \mathbf{x}^\top \left( \mathbf{x} - f(\mathbf{x}) \right)$ the ReD-term.

To solve problem in~\eqref{eq:inverseProblem}, we employ the Alternating Direction Method of Multipliers (ADMM)~\cite{boyd2011distributed}, which enables flexible splitting of the optimization terms. In particular, the Block-Matching and 4D filtering (BM4D) algorithm~\cite{maggioni2012video} is used as the implicit denoiser. The source code is publicly available at: \href{https://github.com/TatianaGelvez/TD-CM-PAM}{https://github.com/TatianaGelvez/TD-CM-PAM}.

After estimating the signal $\hat{\mathbf{x}}$, we construct a beamformed 2D spatial map $\mathbf{X} \in \mathbb{R}^{N_x \times N_z}$ representing the power of the cavitation activity, by summing the squared values for each spatial position across the temporal dimension, i.e., 
\begin{equation}
\mathbf{X}_{i_n,j_n} = \sum \hat{\mathbf{x}}_n^2,
\end{equation}
where the index tuple $(i_n, j_n) \leftrightarrow n$ corresponds to the 2D coordinates associated with the flattened index $n$.

\section{Experiments and Results}
\label{sec:Experiments}
\begin{figure}[t!]
 \centering
 \includegraphics[width=1\linewidth]{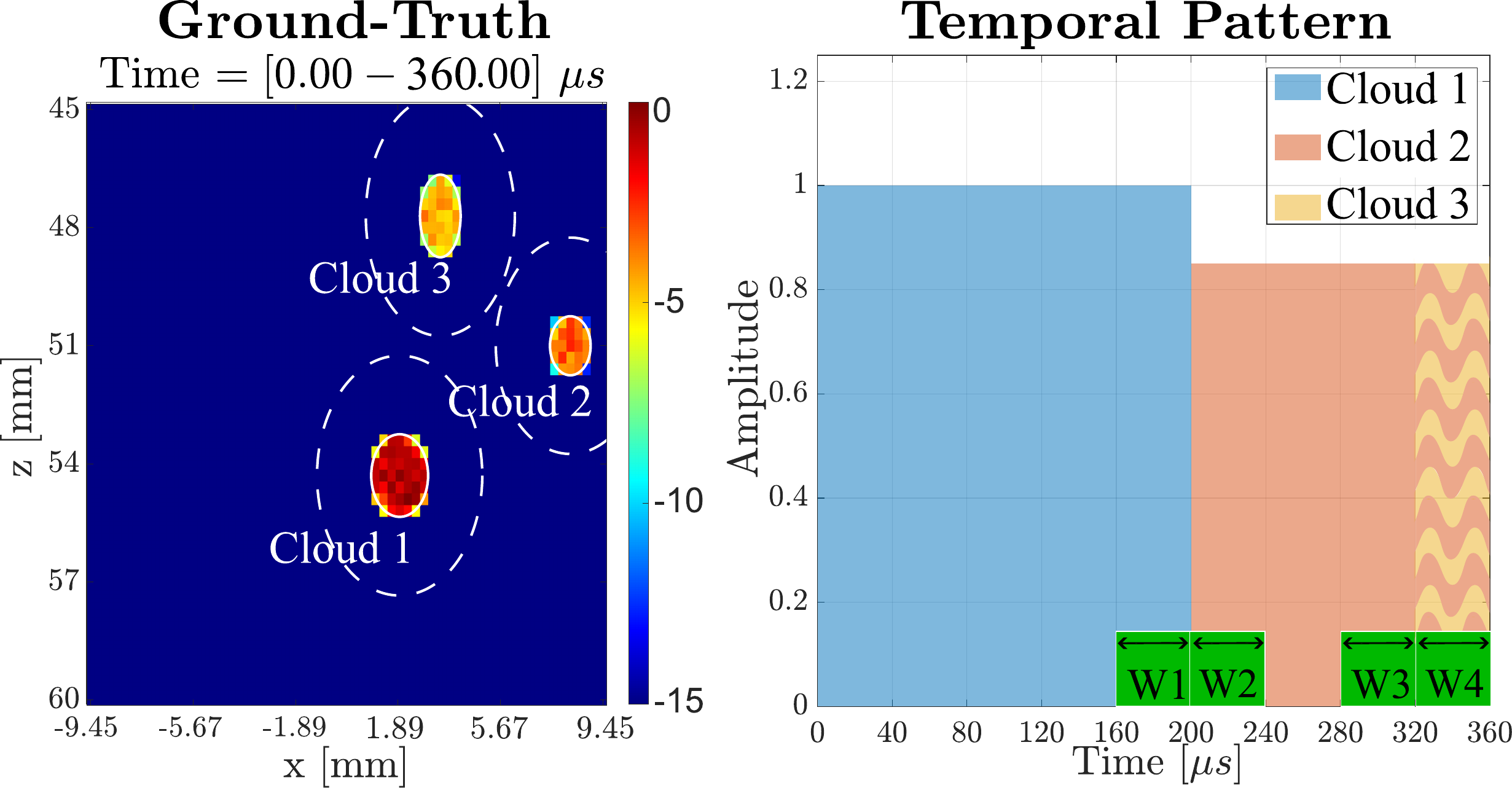}
 \caption{Simulated scenario with three clouds evolving in time.}
 \label{fig:GT}
\end{figure}

This section evaluates the proposed TD-CM-PAM$_\text{SpReD}$ in scenarios with time-evolving cavitation. The experimental setup involves three microbubble clouds whose centers are located at $(-1.95, 54.3)$~mm, $(8.25, 51)$~mm, and $(3.45, 47.7)$~mm in the lateral–axial $(x, 0, z)$ plane. The clouds differ in size, shape, and amplitude, and appear and disappear at different intervals, with partial overlap between the second and third clouds in the last $40~\mu$s, as shown in Figure~\ref{fig:GT}. The RF signals are recorded over $N_t = 360~\mu\text{s}$ using a linear array with $N_m = 64$ sensors, with additive Gaussian noise at 10 dB.

We evaluate the spatial quality using the Contrast-to-Noise Ratio (CNR)~\cite{liebgott2016plane} and the Dice coefficient~\cite{rodriguez2019generalized}. Both metrics are computed as described in~\cite{GelvezBarrera2025}.

\begin{table}[b!]
\centering
\small
\renewcommand{\arraystretch}{1.25}
\caption{Quantitative Performance (Mean (std))}
\begin{tabularx}{\linewidth}{X|c|c}
\hline
\makecell[c]{\textbf{Method}} & \textbf{CNR {[}dB{]}} & \textbf{DICE {[}mm{]}} \\ 
\hline \hline

\textbf{FD-DAS} 
& $1.86\,(0.60)$ 
& $0.09\,(0.13)$ \\ \hline

\textbf{FD-RCB} 
& $0.06\,(1.53)$ 
& $0.19\,(0.22)$ \\ \hline

\textbf{FD-SpTV} 
& $-3.67\,(10.06)$ 
& $0.21\,(0.30)$ \\ \hline

\textbf{TD-DAS} 
& $2.82\,(0.44)$ 
& $0.20\,(0.02)$ \\ \hline

\textbf{TD-LM-PAM$_\text{SpRed}$} 
& $\mathbf{4.20\,(0.57)}$ 
& $\textbf{{0.47\,(0.08)}}$ \\ \hline

\end{tabularx}
\label{tab:CloudResults}
\end{table}

The performance is compared against state-of-the-art beamforming methods in both the time and frequency domains, including TD-DAS~\cite{gyongy2009passive}, FD-DAS~\cite{haworth2016quantitative}, FD-RCB~\cite{lu2018passive}, and FD-CMF-SpTV~\cite{lachambre2024inverse}. Hyperparameters for FD methods follow~\cite{lachambre2024inverse}, with $K = 130$ and $90\%$ overlap for CSM estimation. For the proposed TD-CM-PAM$_\text{SpReD}$, the step size is fixed at $\rho = 0.25$, the sparsity parameter is set to $\lambda = \mathrm{quantile}(|\mathbf{A}^\mathsf{T} \mathbf{y}|,\, 0.95)$, and the denoising regularization parameter is selected via cross-validation.

To identify cavitation cloud locations over time, we construct independent spatial maps with $N_x \times N_z = 64 \times 51$ spatial pixels using non-overlapping windows of $40~\mu\text{s}$, yielding nine spatial maps over the $360~\mu\text{s}$ signal duration. Accordingly, temporal windows of $40~\mu\text{s}$ are used to estimate the spatial maps with TD-based methods. However, for FD-based methods, the minimum theoretical integration interval required to obtain a well-conditioned CSM estimation is $64~\mu\text{s}$, given the selected values of $K$ and the overlap percentage. Based on numerical experiments, we use $200~\mu\text{s}$ windows to estimate the spatial maps and achieve stable FD-based estimation. While such longer integration windows ensure adequate spatial resolution, they may limit temporal tracking capability.

Table~\ref{tab:CloudResults} reports the quantitative performance over the temporal windows. TD-based methods consistently outperform FD-based methods by up to $7$~dB in CNR and $0.6$ in Dice. These results evidence that the FD methods fail to accurately localize microbubble clouds in short temporal windows as they rely on longer intervals, leading to cloud misidentification. 

\begin{figure}[b!]
\centering
\setlength{\tabcolsep}{1pt}
\renewcommand{\arraystretch}{0}
\begin{tabular}{c cccc}

& \multicolumn{4}{c}{\small \textbf{Time interval [$\mu$s]}} \\
& \small $[160 - 200]$ & \small$[200 - 240]$ & \small $[280 - 320]$ & \small $[320 - 360]$ \\

\rotatebox{90}{\parbox{2.1cm}{\centering \textbf{GT}}} &
\begin{tikzpicture}\node[anchor=south west,inner sep=0,draw=magenta,line width=1.25pt]
{\includegraphics[width=0.11\textwidth]{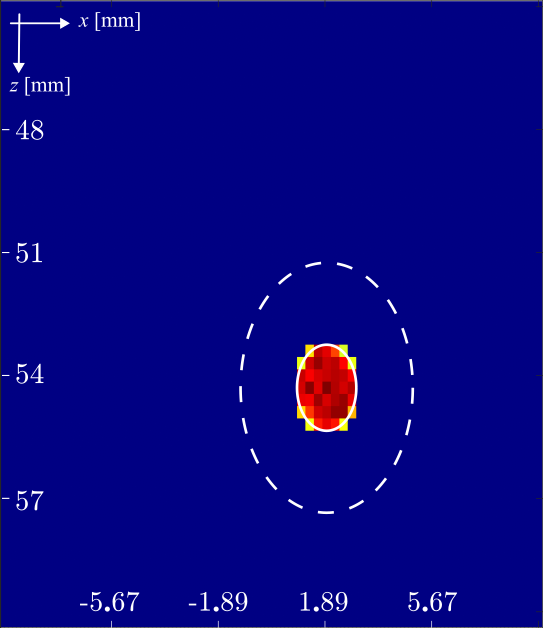}};\end{tikzpicture} &
\begin{tikzpicture}\node[anchor=south west,inner sep=0,draw=magenta,line width=1.25pt]
{\includegraphics[width=0.11\textwidth]{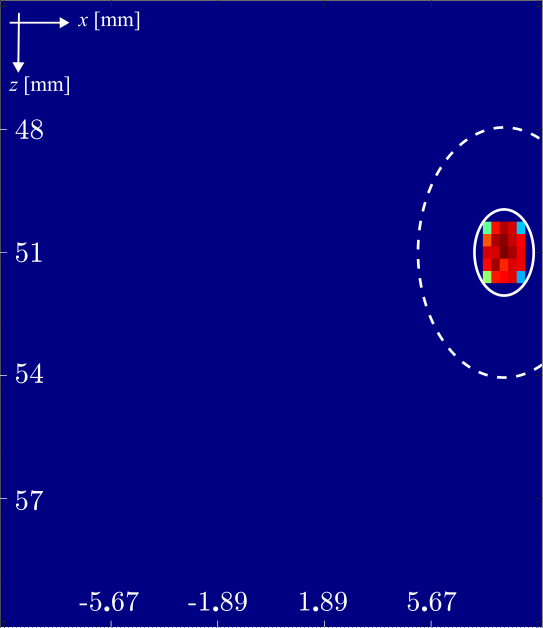}};\end{tikzpicture} &
\begin{tikzpicture}\node[anchor=south west,inner sep=0,draw=magenta,line width=1.25pt]
{\includegraphics[width=0.11\textwidth]{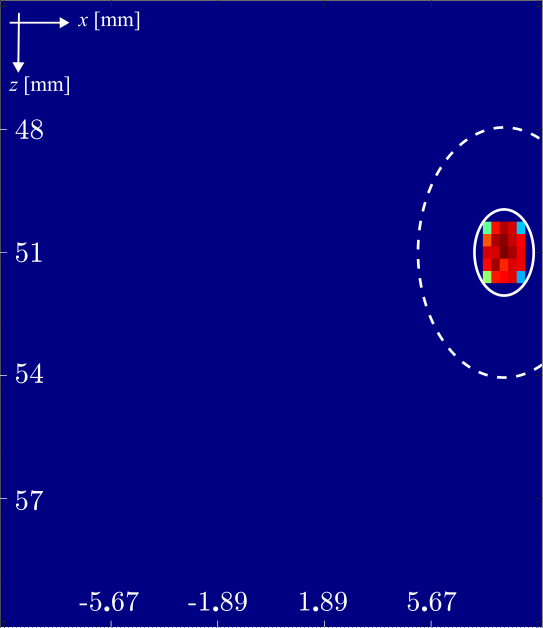}};\end{tikzpicture} &
\begin{tikzpicture}\node[anchor=south west,inner sep=0,draw=magenta,line width=1.25pt]
{\includegraphics[width=0.11\textwidth]{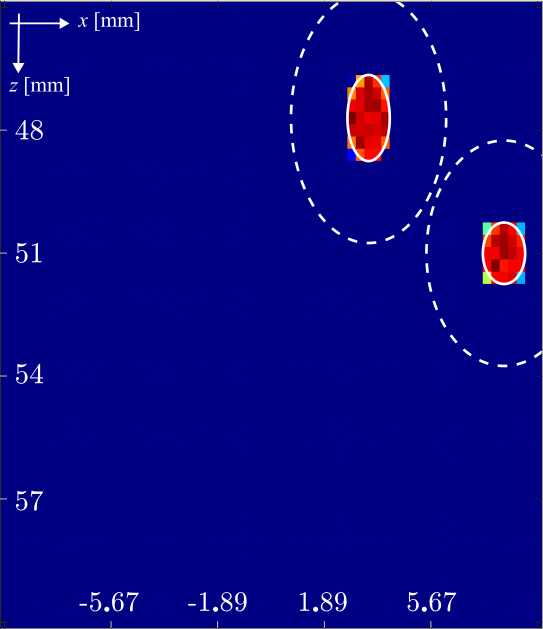}};\end{tikzpicture} \\

\rotatebox{90}{\parbox{2.1cm}{\centering \textbf{FD-DAS}}} &
\begin{tikzpicture}\node[anchor=south west,inner sep=0,draw=magenta,line width=1.25pt]
{\includegraphics[width=0.11\textwidth]{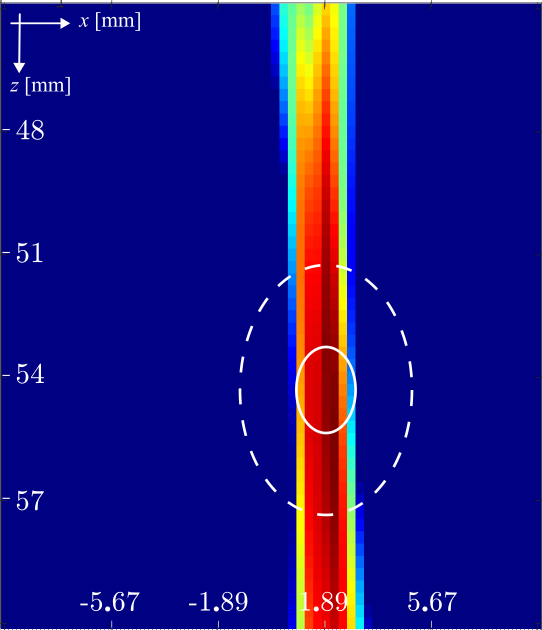}};\end{tikzpicture} &
\begin{tikzpicture}\node[anchor=south west,inner sep=0,draw=magenta,line width=1.25pt]
{\includegraphics[width=0.11\textwidth]{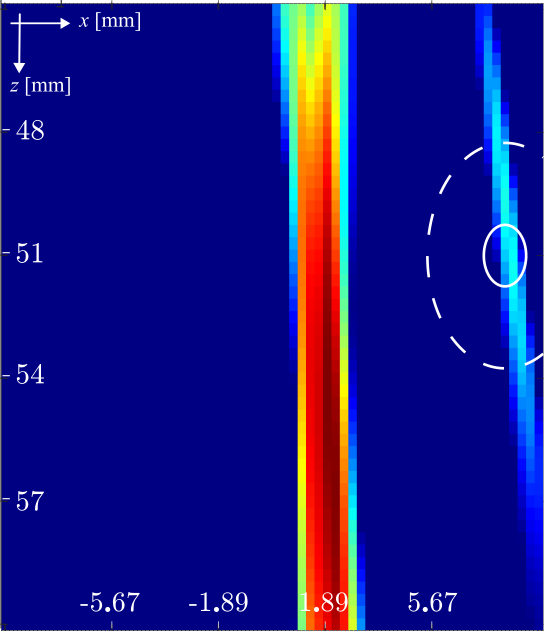}};\end{tikzpicture} &
\begin{tikzpicture}\node[anchor=south west,inner sep=0,draw=magenta,line width=1.25pt]
{\includegraphics[width=0.11\textwidth]{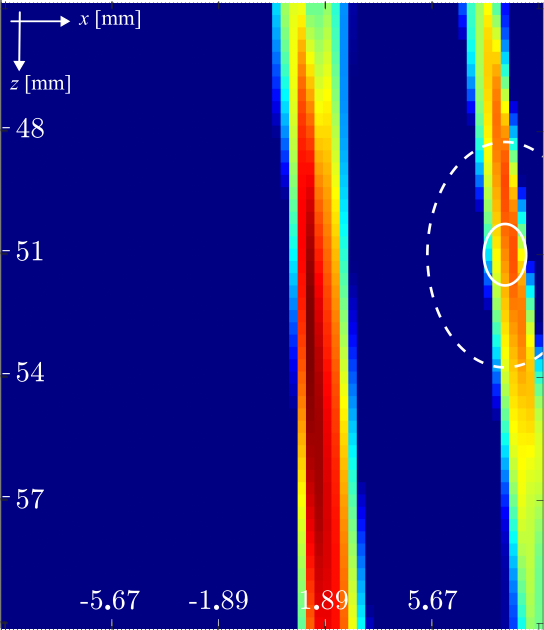}};\end{tikzpicture} &
\begin{tikzpicture}\node[anchor=south west,inner sep=0,draw=magenta,line width=1.25pt]
{\includegraphics[width=0.11\textwidth]{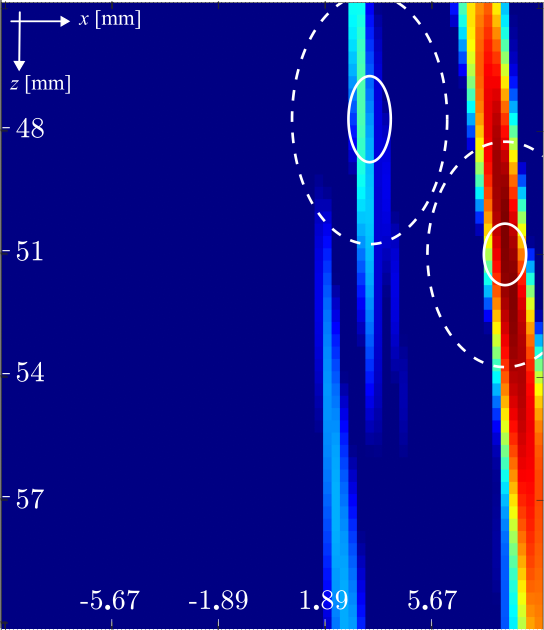}};\end{tikzpicture} \\

\rotatebox{90}{\parbox{2.1cm}{\centering \textbf{FD-RCB}}} &
\begin{tikzpicture}\node[anchor=south west,inner sep=0,draw=magenta,line width=1.25pt]
{\includegraphics[width=0.11\textwidth]{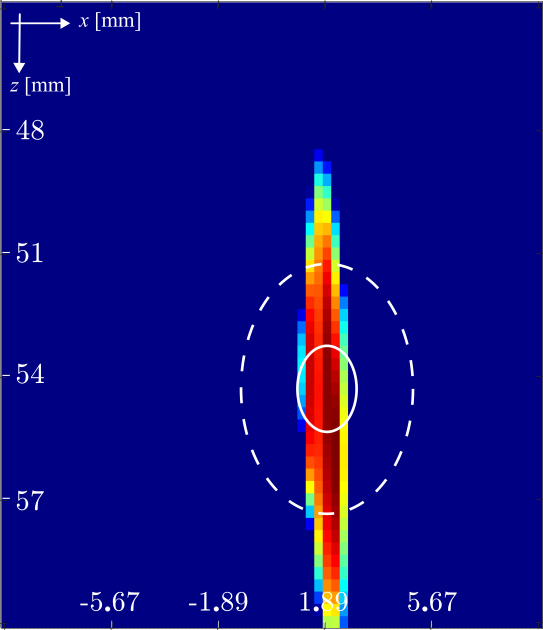}};\end{tikzpicture} &
\begin{tikzpicture}\node[anchor=south west,inner sep=0,draw=magenta,line width=1.25pt]
{\includegraphics[width=0.11\textwidth]{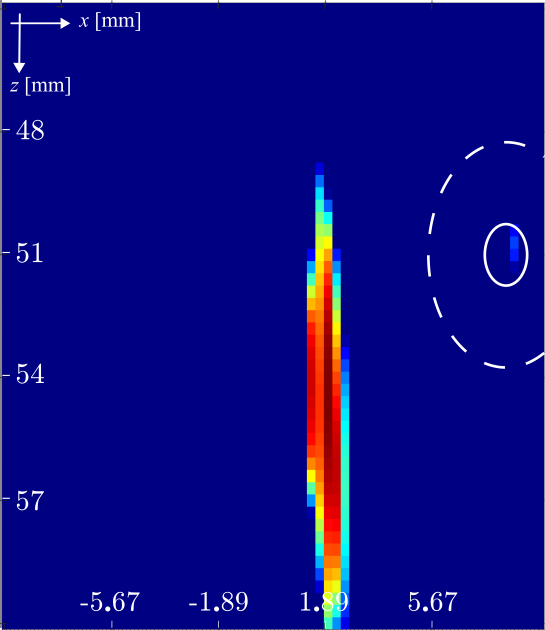}};\end{tikzpicture} &
\begin{tikzpicture}\node[anchor=south west,inner sep=0,draw=magenta,line width=1.25pt]
{\includegraphics[width=0.11\textwidth]{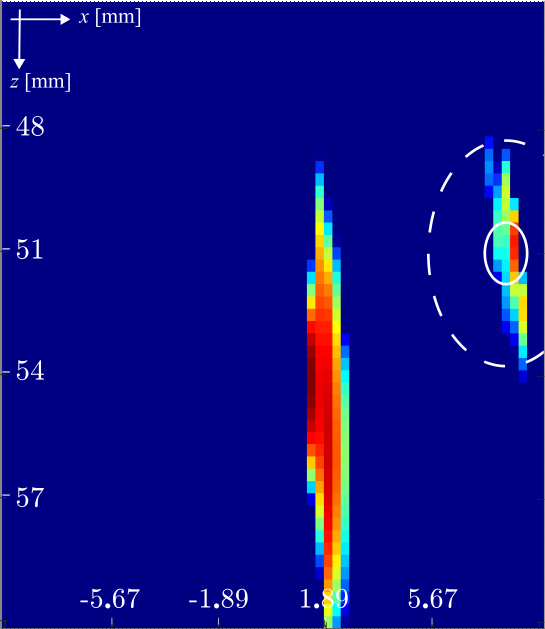}};\end{tikzpicture} &
\begin{tikzpicture}\node[anchor=south west,inner sep=0,draw=magenta,line width=1.25pt]
{\includegraphics[width=0.11\textwidth]{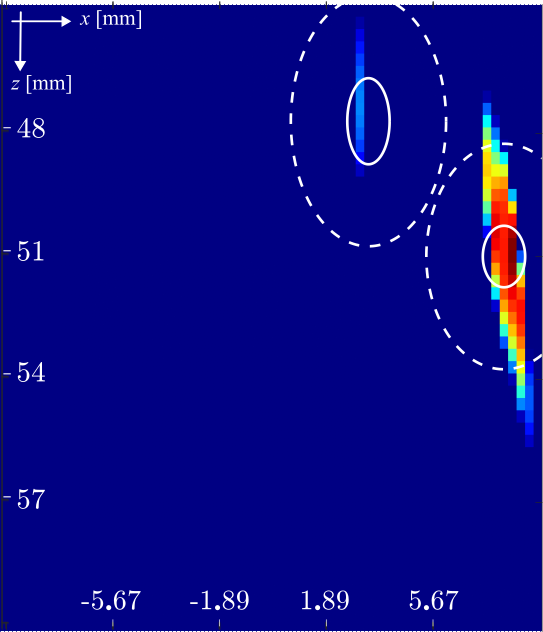}};\end{tikzpicture} \\

\rotatebox{90}{\parbox{2.1cm}{\centering \textbf{FD-spTV}}} &
\begin{tikzpicture}\node[anchor=south west,inner sep=0,draw=magenta,line width=1.25pt]
{\includegraphics[width=0.11\textwidth]{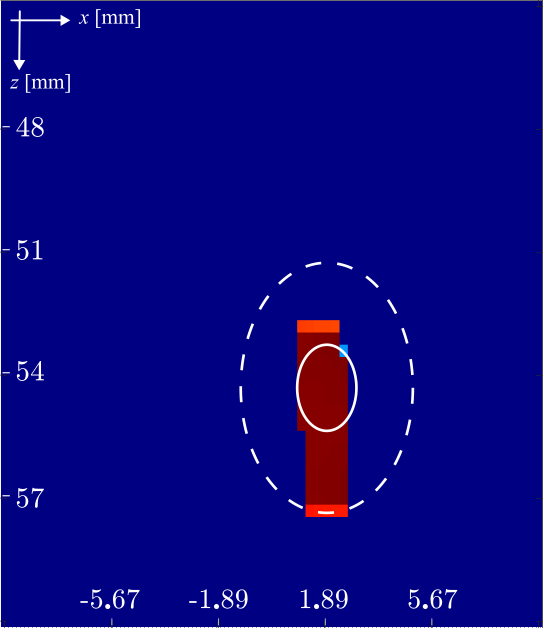}};\end{tikzpicture} &
\begin{tikzpicture}\node[anchor=south west,inner sep=0,draw=magenta,line width=1.25pt]
{\includegraphics[width=0.11\textwidth]{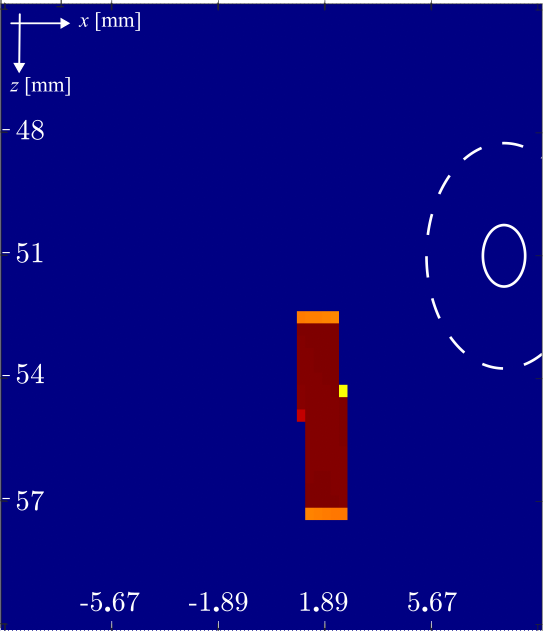}};\end{tikzpicture} &
\begin{tikzpicture}\node[anchor=south west,inner sep=0,draw=magenta,line width=1.25pt]
{\includegraphics[width=0.11\textwidth]{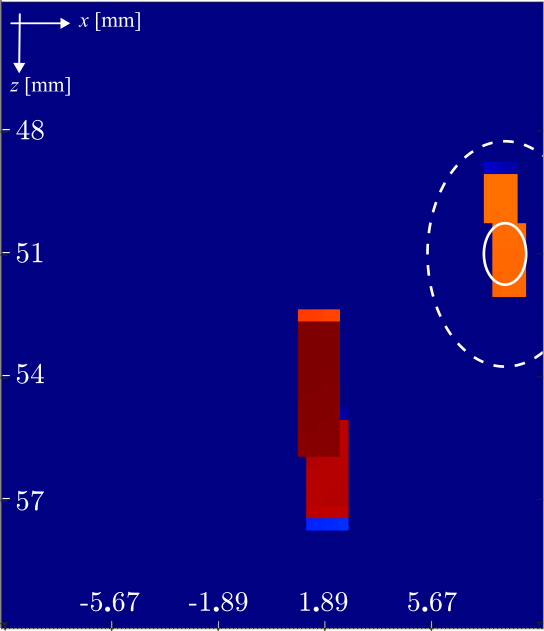}};\end{tikzpicture} &
\begin{tikzpicture}\node[anchor=south west,inner sep=0,draw=magenta,line width=1.25pt]
{\includegraphics[width=0.11\textwidth]{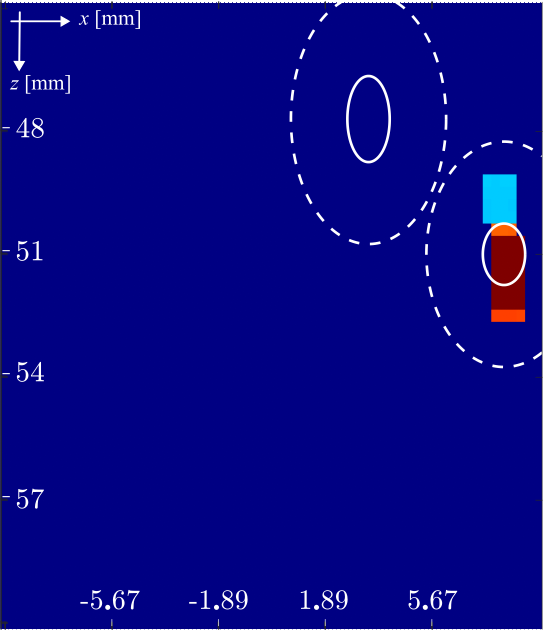}};\end{tikzpicture} \\

\rotatebox{90}{\parbox{2.1cm}{\centering \textbf{TD-DAS}}} &
\begin{tikzpicture}\node[anchor=south west,inner sep=0,draw=magenta,line width=1.25pt]
{\includegraphics[width=0.11\textwidth]{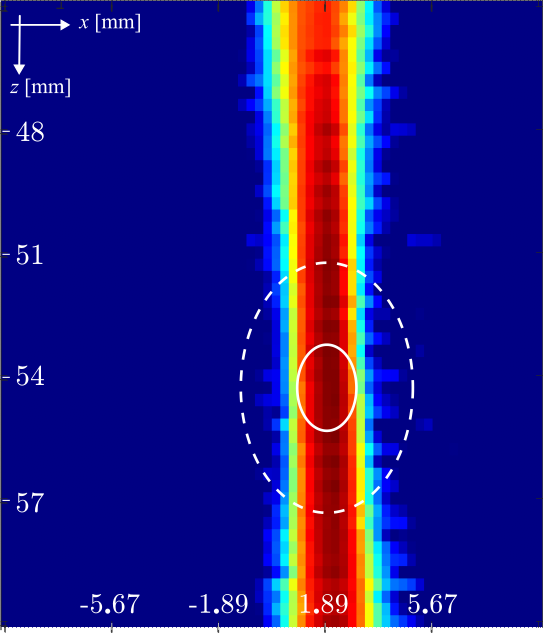}};\end{tikzpicture} &
\begin{tikzpicture}\node[anchor=south west,inner sep=0,draw=magenta,line width=1.25pt]
{\includegraphics[width=0.11\textwidth]{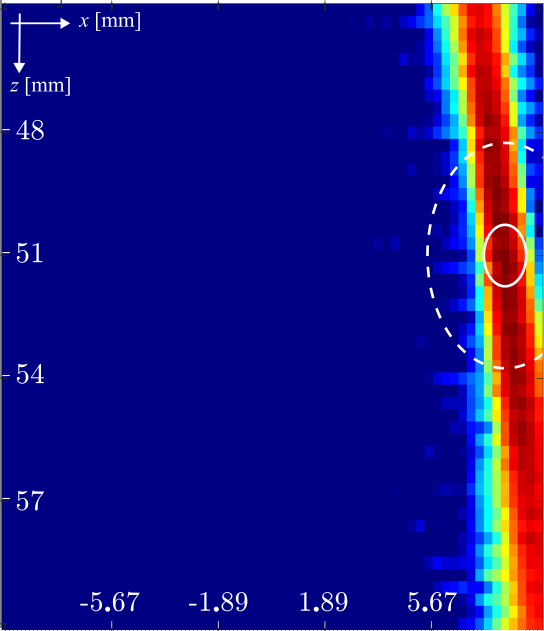}};\end{tikzpicture} &
\begin{tikzpicture}\node[anchor=south west,inner sep=0,draw=magenta,line width=1.25pt]
{\includegraphics[width=0.11\textwidth]{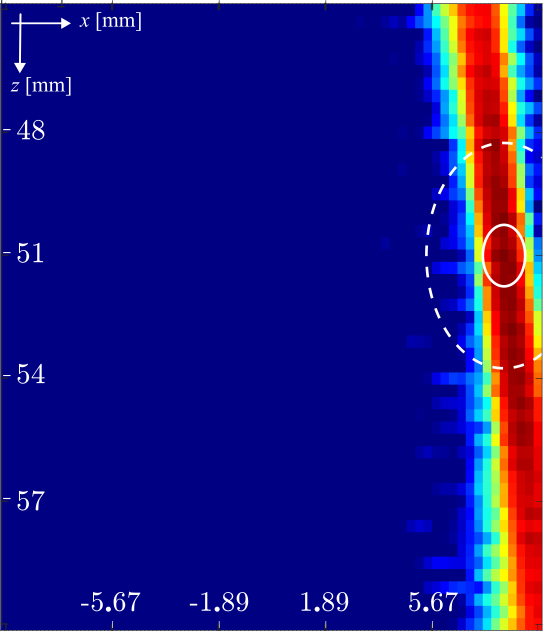}};\end{tikzpicture} &
\begin{tikzpicture}\node[anchor=south west,inner sep=0,draw=magenta,line width=1.25pt]
{\includegraphics[width=0.11\textwidth]{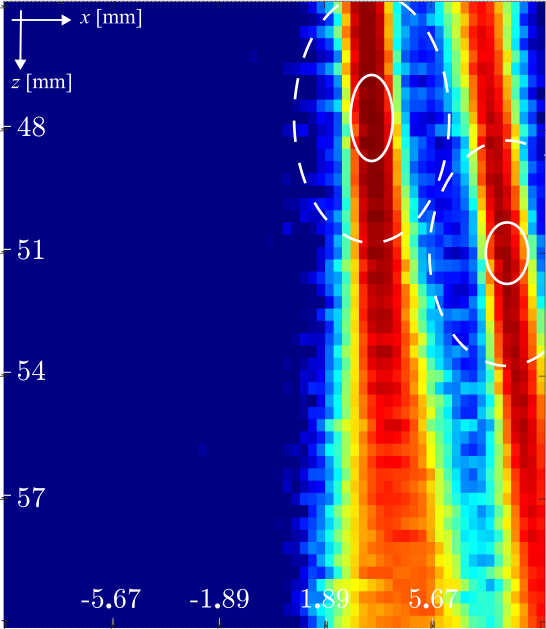}};\end{tikzpicture} \\

\rotatebox{90}{\parbox{2.1cm}{\centering \textbf{Ours}}} &
\begin{tikzpicture}\node[anchor=south west,inner sep=0,draw=magenta,line width=1.25pt]
{\includegraphics[width=0.11\textwidth]{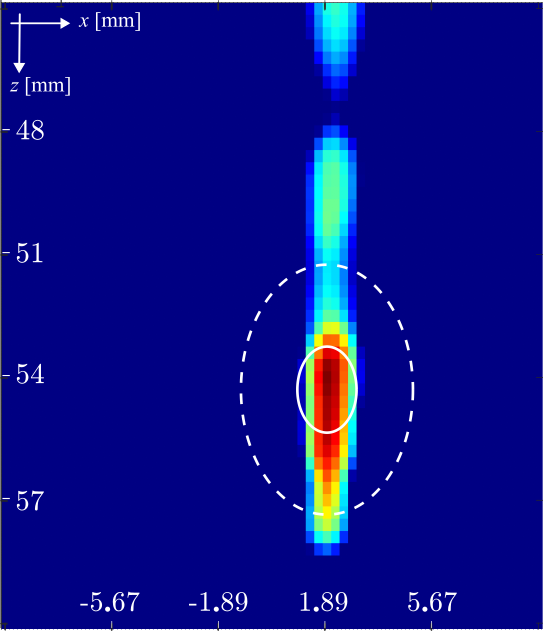}};\end{tikzpicture} &
\begin{tikzpicture}\node[anchor=south west,inner sep=0,draw=magenta,line width=1.25pt]
{\includegraphics[width=0.11\textwidth]{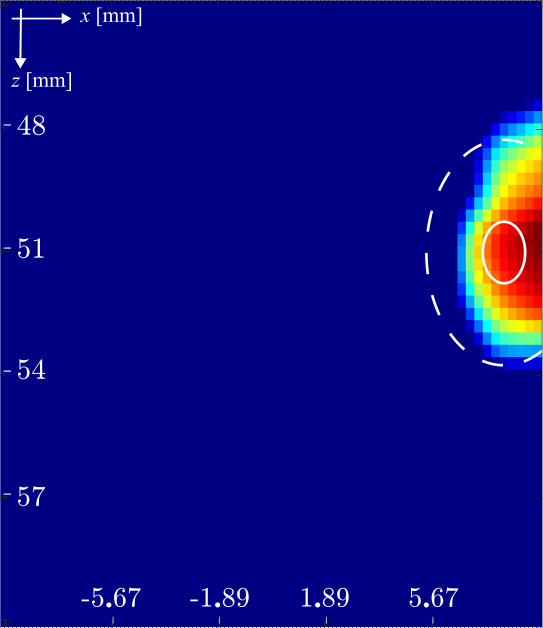}};\end{tikzpicture} &
\begin{tikzpicture}\node[anchor=south west,inner sep=0,draw=magenta,line width=1.25pt]
{\includegraphics[width=0.11\textwidth]{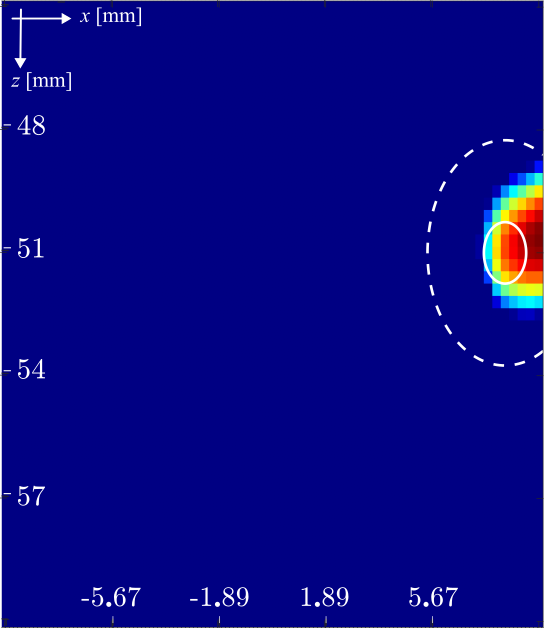}};\end{tikzpicture} &
\begin{tikzpicture}\node[anchor=south west,inner sep=0,draw=magenta,line width=1.25pt]
{\includegraphics[width=0.11\textwidth]{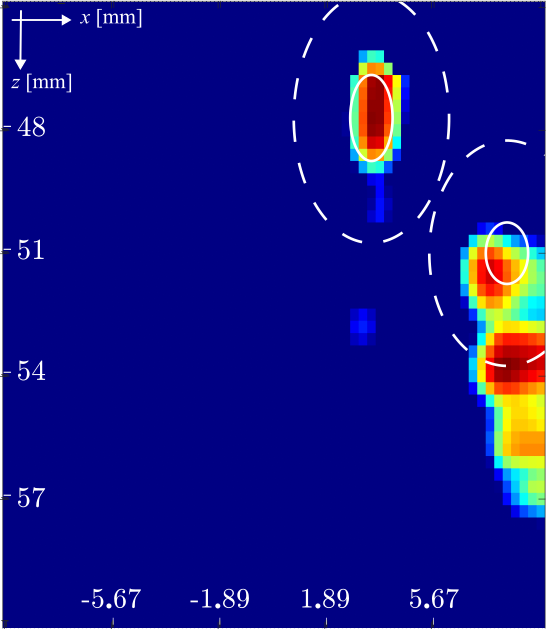}};\end{tikzpicture} \\

\end{tabular}
\caption{Estimated Maps Qualitative comparison.}
\label{fig:QualitativeCom}
\end{figure}

For visual comparison, Fig.~\ref{fig:QualitativeCom} shows four temporal windows, labeled as W$_w$ for $w=1,\dots,4$ in Fig.~\ref{fig:GT}. W$_1$ captures a stationary regime, where the first cloud has persisted over an extended period. In this scenario, all methods perform well, as the cloud is stable and easy to detect. In W$_2$, a second cloud emerges while the first cloud has disappeared. TD-based methods accurately locate the new cloud due to their sensitivity to rapid temporal changes, whereas FD-based methods fail, as their longer integration window continues to emphasize the first cloud. In W$_3$, the second cloud has persisted over time, hence, both methods detect it; however, FD methods still display remnants of the first cloud, reflecting their slower temporal response. Finally, W$_4$ presents the most challenging case, with partially overlapping clouds. Most FD methods fail to detect the new cloud and continue showing the earlier one, while the proposed approach identifies the current cloud and achieves a good reconstruction of cloud shapes.

 \section{Conclusions}
This paper introduced a time-domain convolutional model-based framework for passive beamforming, illustrated through passive acoustic mapping for cavitation monitoring. By reformulating the forward model as a convolution, the proposed approach significantly reduces computational complexity while preserving high temporal resolution and physical interpretability. Combined with a regularized inversion strategy, the method outperforms state-of-the-art time approaches in scenarios with cavitation activity evolving over time. Although demonstrated for passive cavitation, the proposed formulation is general and can be adapted to a wide range of applications involving passive beamforming and spatiotemporal source localization.

\bibliographystyle{IEEEtran} 
\bibliography{references} 

@inproceedings{sivadon2020pisarenko,
  title={Pisarenko class beamformer applied to passive acoustic mapping of ultrasound cavitation},
  author={Sivadon, Audrey and Polichetti, Maxime and B{\'e}ra, Jean-Christophe and Varray, Fran{\c{c}}ois and Nicolas, Barbara and Gilles, Bruno},
  booktitle={Forum Acusticum},
  pages={1061--1064},
  year={2020}
}

@article{polichetti2020use,
  title={Use of the cross-spectral density matrix for enhanced passive ultrasound imaging of cavitation},
  author={Polichetti, Maxime and Varray, Fran{\c{c}}ois and Gilles, Bruno and B{\'e}ra, Jean-Christophe and Nicolas, Barbara},
  journal={IEEE Transactions on Ultrasonics, Ferroelectrics, and Frequency Control},
  volume={68},
  number={4},
  pages={910--925},
  year={2020},
  publisher={IEEE}
}

@article{lu2018passive,
  title={Passive acoustic mapping of cavitation using eigenspace-based robust Capon beamformer in ultrasound therapy},
  author={Lu, Shukuan and Hu, Hong and Yu, Xianbo and Long, Jiangying and Jing, Bowen and Zong, Yujin and Wan, Mingxi},
  journal={Ultrasonics sonochemistry},
  volume={41},
  pages={670--679},
  year={2018},
  publisher={Elsevier}
}

@article{haworth2016quantitative,
  title={Quantitative frequency-domain passive cavitation imaging},
  author={Haworth, Kevin J and Bader, Kenneth B and Rich, Kyle T and Holland, Christy K and Mast, T Douglas},
  journal={IEEE transactions on ultrasonics, ferroelectrics, and frequency control},
  volume={64},
  number={1},
  pages={177--191},
  year={2016},
  publisher={IEEE}
}

@article{maggioni2012video,
  title={Video denoising, deblocking, and enhancement through separable 4-D nonlocal spatiotemporal transforms},
  author={Maggioni, Matteo and Boracchi, Giacomo and Foi, Alessandro and Egiazarian, Karen},
  journal={IEEE Transactions on image processing},
  volume={21},
  number={9},
  pages={3952--3966},
  year={2012},
  publisher={IEEE}
}

@article{boyd2011distributed,
  title={Distributed optimization and statistical learning via the alternating direction method of multipliers},
  author={Boyd, Stephen and Parikh, Neal and Chu, Eric and Peleato, Borja and Eckstein, Jonathan and others},
  journal={Foundations and Trends{\textregistered} in Machine learning},
  volume={3},
  number={1},
  pages={1--122},
  year={2011},
  publisher={Now Publishers, Inc.}
}

@article{gyongy2009passive,
  title={Passive spatial mapping of inertial cavitation during HIFU exposure},
  author={Gyongy, Miklos and Coussios, Constantin-C},
  journal={IEEE Transactions on Biomedical Engineering},
  volume={57},
  number={1},
  pages={48--56},
  year={2009},
  publisher={IEEE}
}

@article{boulos2018weighting,
  title={Weighting the passive acoustic mapping technique with the phase coherence factor for passive ultrasound imaging of ultrasound-induced cavitation},
  author={Boulos, Paul and Varray, Fran{\c{c}}ois and Poizat, Adrien and Ramalli, Alessandro and Gilles, Bruno and Bera, Jean-Christophe and Cachard, Christian},
  journal={IEEE transactions on ultrasonics, ferroelectrics, and frequency control},
  volume={65},
  number={12},
  pages={2301--2310},
  year={2018},
  publisher={IEEE}
}

@article{coussios2008applications,
  title={Applications of acoustics and cavitation to noninvasive therapy and drug delivery},
  author={Coussios, Constantin C and Roy, Ronald A},
  journal={Annu. Rev. Fluid Mech.},
  volume={40},
  number={1},
  pages={395--420},
  year={2008},
  publisher={Annual Reviews}
}

@article{crake2018passive,
  title={Passive acoustic mapping and B-mode ultrasound imaging utilizing compressed sensing for real-time monitoring of cavitation-enhanced drug delivery},
  author={Crake, Calum and Finn, Se{\'a}n and Marsac, Laurent and Gray, Michael and Carlisle, Robert and Coussios, Constantin and Coviello, Christian},
  journal={The Journal of the Acoustical Society of America},
  volume={143},
  number={3\_Supplement},
  pages={1872--1872},
  year={2018},
  publisher={Acoustical Society of America}
}

@article{coviello2015passive,
  title={Passive acoustic mapping utilizing optimal beamforming in ultrasound therapy monitoring},
  author={Coviello, Christian and Kozick, Richard and Choi, James and Gy{\"o}ngy, Mikl{\'o}s and Jensen, Carl and Smith, Penny Probert and Coussios, Constantin-C},
  journal={The Journal of the Acoustical Society of America},
  volume={137},
  number={5},
  pages={2573--2585},
  year={2015},
  publisher={AIP Publishing}
}

@article{lachambre2024inverse,
  title={An inverse method using Cross spectral Matrix Fitting for passive cavitation imaging},
  author={Lachambre, C{\'e}lestine and Basarab, Adrian and B{\'e}ra, Jean-Christophe and Nicolas, Barbara and Varray, Fran{\c{c}}ois and Gilles, Bruno},
  journal={IEEE Transactions on Ultrasonics, Ferroelectrics, and Frequency Control},
  year={2024},
  publisher={IEEE}
}

@article{saletesresearch,
  title={Research Article In Vitro Demonstration of Focused Ultrasound Thrombolysis Using Bifrequency Excitation},
  author={Saletes, Izella and Gilles, Bruno and Auboiroux, Vincent and Bendridi, Nadia and Salomir, Rares and B{\'e}ra, Jean-Christophe},
  journal={BioMed Research International},
  volume={2014},
  pages={1--10},
  year={2014},
  publisher={Hindawi Publishing Corporation},
  doi={10.1155/2014/518787},
  url={http://dx.doi.org/10.1155/2014/518787}
}

@article{tan2024modelling,
  title={Modelling the dynamics of microbubble undergoing stable and inertial cavitation: Delineating the effects of ultrasound and microbubble parameters on sonothrombolysis},
  author={Tan, Zhi Qi and Ooi, Ean Hin and Chiew, Yeong Shiong and Foo, Ji Jinn and Ng, Yin Kwee and Ooi, Ean Tat},
  journal={Biocybernetics and Biomedical Engineering},
  volume={44},
  number={2},
  pages={358--368},
  year={2024},
  publisher={Elsevier}
}

@article{denner2023modeling,
  title={Modeling acoustic emissions and shock formation of cavitation bubbles},
  author={Denner, Fabian and Schenke, S{\"o}ren},
  journal={Physics of Fluids},
  volume={35},
  number={1},
  year={2023},
  publisher={AIP Publishing}
}

@article{moonen2025focused,
  title={Focused Ultrasound: Noninvasive Image-Guided Therapy},
  author={Moonen, Chrit TW and Kilroy, Joseph P and Klibanov, Alexander L},
  journal={Investigative Radiology},
  volume={60},
  number={3},
  pages={205--219},
  year={2025},
  publisher={LWW}
}

@inproceedings{liebgott2016plane,
  title={Plane-wave imaging challenge in medical ultrasound},
  author={Liebgott, Herve and Rodriguez-Molares, A and Cervenansky, F and Jensen, J{\o}rgen Arendt and Bernard, Olivier},
  booktitle={2016 IEEE International ultrasonics symposium (IUS)},
  pages={1--4},
  year={2016},
  organization={IEEE}
}

@article{chen2016dynamic,
  title={Dynamic behavior of microbubbles during long ultrasound tone-burst excitation: mechanistic insights into ultrasound-microbubble mediated therapeutics using high-speed imaging and cavitation detection},
  author={Chen, Xucai and Wang, Jianjun and Pacella, John J and Villanueva, Flordeliza S},
  journal={Ultrasound in medicine \& biology},
  volume={42},
  number={2},
  pages={528--538},
  year={2016},
  publisher={Elsevier}
}

@misc{GelvezBarrera2025,
  title         = {Time-Domain Linear Model-based Framework for Passive Acoustic Mapping of Cavitation Activity},
  author        = {Gelvez-Barrera, Tatiana and Nicolas, Barbara and Kouam{\'e}, Denis and Gilles, Bruno and Basarab, Adrian},
  year          = {2025},
  eprint        = {2511.20551},
  archivePrefix = {arXiv},
  primaryClass  = {eess.SP},
  doi           = {10.48550/arXiv.2511.20551},
  url           = {https://arxiv.org/abs/2511.20551}
}

@article{rodriguez2019generalized,
  title={The generalized contrast-to-noise ratio: A formal definition for lesion detectability},
  author={Rodriguez-Molares, Alfonso and Rindal, Ole Marius Hoel and D’hooge, Jan and M{\aa}s{\o}y, Svein-Erik and Austeng, Andreas and Bell, Muyinatu A Lediju and Torp, Hans},
  journal={IEEE transactions on ultrasonics, ferroelectrics, and frequency control},
  volume={67},
  number={4},
  pages={745--759},
  year={2019},
  publisher={IEEE}
}

\end{document}